\documentstyle[aps,psfig,amssymb]{revtex}

\newcommand{\Lcsb}{\Lambda_{\mbox {\tiny CSB}}}

\begin{document}

\title{Meson exchange and nucleon polarizabilities in the quark model}

\author{F. Cano$^1$ and  M. Traini$^{1,2}$}

\address{$^1$Dipartimento di Fisica, Universit\`a degli Studi di Trento \\
I-38050 Povo, Italy \\
$^2$ Istituto Nazionale di Fisica Nucleare, G.C. Trento.}
\date{\today}
\maketitle

\begin{abstract}
Modifications to the nucleon electric polarizability 
induced by pion and sigma exchange in the $q-q$ potentials
are studied by means of sum rule techniques within a 
non-relativistic quark model. Contributions from meson exchange
interactions are
found to be small and in general reduce the quark core 
polarizability for a number of hybrid and one-boson-exchange $q-q$
models. These results can be explained by the constraints that the
baryonic spectrum impose on the short
range behavior of the mesonic interactions. 
\end{abstract}


\section{Introduction}

Electric (${\alpha}$) and magnetic (${\beta}$) polarizabilities 
are fundamental observables characterizing the response of the nucleon to 
an external (quasi)-static electromagnetic field. In particular ${\alpha}$ controls the 
deformation induced by the electric field and can be 
experimentally determined for the proton by measuring 
the Compton polarizability $\bar {\alpha}$ \cite{FEDERSPIEL91}.
In a non-relativistic approach $\bar {\alpha}$ differs from $\alpha$
by a so called retardation term $\Delta \alpha$:

\begin{equation}
\bar{\alpha} = \alpha + \Delta \alpha \; , 
\end{equation}

\noindent
where 

\begin{equation}
\Delta \alpha = \frac{e}{3 M} \langle 0 | \sum_{i=1}^3 e_i (\vec{r}_i - 
\vec{R})^2|0 \rangle  = \frac{e^2 \langle r^2 \rangle_{ch} }{3 M} \; ,
\end{equation}

\noindent and the (static) polarizability, obtained from the
low-energy theorems \cite{ERICSON73} is

\begin{equation}
\label{dipbasic}
\alpha = 2 \sum_{n \neq 0} \frac{|\langle n|D_z |0 \rangle|^2 }{E_n - E_0}\,\,.
\end{equation}

In the previous expressions $|0 \rangle$ is the ground state 
(nucleon) and $|n\rangle$ the excited states allowed by the dipole operator 

\begin{equation}
D_z = \sum_{i=1}^3 e_i (\vec{r}_i - \vec{R})_z\,\,. 
\label{dipoperator}
\end{equation}

\noindent $E_n$ and $E_0 \equiv M$ denote the masses of these excited states 
and the proton respectively, $e_i$ and $\vec{r}_i$ stand for the charge and 
position of the (constituent) quarks, and $\vec{R}$ is the center of mass 
coordinate.

A recent analysis of Compton scattering experiments \cite{MACGIBBON95} yields

\begin{equation}
\bar{\alpha}^p = \alpha^p + \Delta \alpha^p = 
(12.1 \pm 0.8 \pm 0.5) \; 10^{-4} \; \mbox {fm}^{3}\,\,,
\label{alphapexp}
\end{equation}

	The static polarizability of the proton (${\alpha}^p$) is
obtained by subtracting the retardation contribution in
(\ref{alphapexp}), while for the neutron one has access to $\alpha^n$
directly \cite{polariz97} by means of a neutron-Nucleus scattering (specifically n-Pb
\cite{SCHMI91}) and the final results read
\cite{FEDERSPIEL91,SCHMI91}:

\begin{equation}
\label{alphaexp}
\begin{array}{rcl}
{\alpha}^p & = & (7.0 \pm 2.2 \pm 1.3) \; 10^{-4} \; \mbox {fm}^{3} \; ,\\
{\alpha}^n & = & (12.0 \pm 1.5 \pm 2.0) \; 10^{-4} \; \mbox  {fm}^{3}.
\end{array}
\end{equation}

	These observables have been calculated in a wide variety of hadronic 
models, including bag and soliton models, chiral perturbation theory and 
dispersion relation methods (see \cite{LVOV93} for a review). The
non-relativistic constituent quark model seems to present a serious and 
peculiar problem \cite{SCHOBERL86,DESANCTIS90,LUCHA90,TRAINI94}: it
cannot reproduce 
the spectrum and the experimental values
(\ref{alphaexp}) simultaneously. 
This can be straightforwardly illustrated within  the harmonic oscillator 
model \cite{HOLSTEIN94}. In this case the sum over excited states can be 
performed analytically and one gets:

\begin{equation}
\alpha^p = \alpha^n = \frac{2}{9}e^2 M \langle r_p^2 \rangle^2_{ch}=
2\,{e^2 \over M}\,{1\over (\hbar \omega)^2} \; .
\label{alphaho}
\end{equation} 

	Eq.(\ref{alphaho}) largely underestimates the proton polarizability 
under the requirement of a small charge radius as suggested by spectroscopy
($\hbar\omega \approx 600$ MeV). The same conclusion holds  
within a large class of more realistic potentials \cite{BIASIOLI99}.
Moreover, the (approximate) charge symmetry exhibited by
non-relativistic quark models makes 
difficult to explain the experimental difference $\alpha^n-\alpha^p$. It is
intuitive to argue that the meson cloud surrounding the core of quarks
would substantially contribute to the total polarizability of the
nucleon \cite{LVOV93,SCHOBERL86}. As a matter of fact, it has been
shown \cite{WEINER85} that in chiral quark 
models a large fraction of $\alpha$ comes from the coupling of the 
photon to the charged pion fields, i.e. from the cloud. 
Though the mesonic cloud seems to have little effect on the baryon
mass calculation, it is 
essential to understand many electromagnetic properties of the baryons, 
the most immediate example being the charge radius of the proton.

	Recently some quark models have emerged that consider mesonic
exchanges in the $q-q$ potential
\cite{FERNANDEZ93,BUCHMANN94,GLOZMAN96}, as a phenomenological way to
include chiral symmetry breaking.  In addition a large debate rose on
the need to invoke Goldston boson exchange in connection with the
notion of constituent quark \cite{ISGUR99}, or the possibility of
reproducing the observed spectrum within the simple effective (one)
gluon exchange (OGE) model \cite{GLOZMAN99}.

The consideration of the meson exchange in the potential  
relates two different 'worlds' of degrees of freedom: the 
low-energy nuclear physics degrees of freedom (mesons and baryons) and 
the fundamental description in terms of quarks and gluons. 
At the moment there is
no clear evidence of a smooth transition between these two 
'worlds' \cite{SMOOTH} and a study of the effects induced by 
the $q-q$ meson interaction can shed some light on the problem. 

	In particular one can rise the question: to which extent can
these mesonic degrees of freedom contribute to explain the
electromagnetic properties of the baryon, i.e. relegate the missing
mesonic cloud to play a marginal role. The answer to this question
would be of some interest to understand if OBE-based quark models are
in a better shape than OGE-based (or hybrid) models to describe the
electromagnetic structure of baryons.  
The electric polarizability
is a peculiar observable sensitive, in principle, to both
quark and meson degrees of freedom and its study by means of sum rules
can represent a useful tool to elucidate the comparative advantages
and disadvantages of OBE models and OGE (or hybrid) models.
Furthermore, the sum-rule
method is sensitive to very basic properties of the models, averaging
many fine details of the interaction. 

\section{The theoretical framework}

	Our starting point to calculate the electric static
polarizability, including the effects of the meson and gluon $q-q$
interactions, has to be simple enough to have a direct relation with
the potential model and, in addition, independent on the space of
states introduced by Eq.(\ref{dipbasic}). In the harmonic oscillator
case the sum over excited states can be directly
evaluated but in more realistic scenarios one has to resort to
other techniques such as variational methods \cite{SCHOBERL86} or sum
rules \cite{TRAINI94}\footnote{One could also note that variational
and sum rule approaches are intimately connected
\cite{BIASIOLI99,TRAINIejp96}.}. We make use of a sum rule technique
which, washing out all the complications of the baryonic spectrum,
requires the knowledge of the nucleon wave function only, and
constrains the numerical result to satisfy the stringent inequality
\cite{BIASIOLI99}:

\begin{equation}
\label{alphabounds}
2 \frac{m_0^2(D_z)}{m_1(D_z)} \leq \alpha \leq 2 \frac{m_0(D_z)}{E_{10}},
\end{equation}
 
\noindent 
where $E_{10}$ is the energy gap between the nucleon and the first electric 
dipole excitation, $D_{13}(1520)$, and the moments (sum rules) of the 
dipole operator read

\begin{eqnarray}
\label{m_not}
m_0(D_z) & = & \langle 0 | D_z D_z |0 \rangle \; ,\\
\label{m_one}
m_1(D_z) & = & {1\over 2}\,\langle 0 | [D_z, [H,D_z]] |0 \rangle. 
\end{eqnarray}

	The potential model enters in two basic ways: i) explicitly in the
Hamiltonian, Eq.(\ref{m_one}); ii) implicitly in the ground 
state wave function $|0\rangle$ ($H\,|0\rangle = E_0\,|0\rangle$). 

	The general structure of the potential models we consider
can be written
\begin{equation}
\label{vqqfull}
V_{qq} = V_{\mbox {\scriptsize Conf}} + V_{\mbox {\scriptsize OGE}} +
V_{\mbox {\scriptsize OPE}} + V_{\mbox {\scriptsize OSE}}\,\,,
\end{equation}

\noindent
where $V_{\mbox {\scriptsize Conf}}$ defines the confining 
potential, $V_{\mbox {\scriptsize OGE}}$ embodies the one-gluon 
exchange interaction and $V_{\mbox {\scriptsize OPE}}$ 
($V_{\mbox {\scriptsize OSE}}$) is the one-pion (one-sigma) 
exchange term. The general structure of these interactions is:

\begin{eqnarray}
\label{voge}
V_{\mbox {\scriptsize OGE}}(r_{12}) & = &  \left[ 
V^C_{\mbox {\scriptsize OGE}}(r_{12}) + 
V^S_{\mbox {\scriptsize OGE}}(r_{12}) \vec{\sigma}_1 \cdot
\vec{\sigma}_2 +   V^T_{\mbox {\scriptsize OGE}}(r_{12})
\hat{S}_{12}\right]\\
\label{vope}
V_{\mbox {\scriptsize OPE}}(r_{12}) & = & \left[ 
V^S_{\mbox {\scriptsize OPE}}(r_{12})
\vec{\sigma}_1 \cdot
\vec{\sigma}_2 +   V^T_{\mbox {\scriptsize OPE}}(r_{12})
\hat{S}_{12}\right] \vec{\tau}_i \cdot
\vec{\tau}_j \\
V_{\mbox {\scriptsize OSE}}(r_{12}) & = &
V^C_{\mbox {\scriptsize OSE}}(r_{12})\,\,,
\label{vose}
\end{eqnarray}

\noindent
with $\hat{S}_{ij} = 3 (\vec{\sigma}_i \cdot \hat{r}_{ij}) (\vec{\sigma}_j 
\cdot \hat{r}_{ij}) - (\vec{\sigma}_i \cdot \vec{\sigma}_j)$; $r_{ij}$ is 
the interquark distance and $\vec{\sigma}_i$ ($\vec{\tau}_i$) are the spin 
(isospin) operator of the $i$-th quark.

In order to illustrate the relevance of the meson exchange effects, let us
first calculate the sum rules $m_1(D_z)$ and $m_0(D_z)$ neglecting,
in the Hamiltonian, the contribution due to $V_{\mbox {\scriptsize OPE}}$  and 
$V_{\mbox {\scriptsize OSE}}$, Eqs. (\ref{vope},\ref{vose}). One obtains:

\begin{eqnarray}
\label{simplem0}
m_0(D_z) & = & \frac{1}{3} e^2 \langle r_p^2 \rangle_{ch} + 
\frac{2}{3} e^2 \langle r_n^2 \rangle_{ch} \\
m_1(D_z) & = & \frac{e^2}{3 m_q}
\label{simplem1}
\end{eqnarray}
 
\noindent
for both proton and neutron. The non vanishing value of $m_1(D_z)$ in 
(\ref{simplem1}) is due to the conmutator of the kinetic energy and 
the dipole operator and no additional contribution comes from the 
confining and OGE potentials which commute with the dipole operator
(\ref{dipoperator}). 

	The reference model that we will use throughout the paper will
be the one of Valcarce {\it et al.} \cite{FERNANDEZ93} that contains all
the terms (\ref{voge}-\ref{vose}). The constituent quark
mass is taken $m_q =  M_N/3 = 313$ MeV, and the resulting charge radii of 
the nucleon in this model are 
($\langle r_p^2 \rangle_{ch} = 0.252$ fm$^2$, 
$\langle r_n^2 \rangle_{ch} =  -2.58 \cdot 10^{-2}$ fm$^2$), so
that one obtains, from (\ref{simplem0}-\ref{simplem1}),  

\begin{equation}
\label{cqmestimate}
3.10 \cdot 10^{-4} \; \mbox{fm}^3 \leq \alpha^{p,n} 
\leq 3.76 \cdot 10^{-4} \; \mbox{fm}^3\,\, ,
\end{equation}

\noindent
a result which can be regarded as the typical outcome of a CQM including OGE 
potential, since the $m_q$ and $\langle r^2 \rangle$ used above are
quite common in a large class of models that reproduce the basic
features of the hadronic spectrum (see \cite{BIASIOLI99} for an
extensive study of polarizabilities in a number of quark models).

The Isgur-Karl (IK) model  \cite{ISGURKARL} deserves a specific
comment since the inclusion of  
an unknown potential term to remove the harmonic oscillator degeneracy 
prevents explicit calculations of the 
sum rules.
Nevertheless, since the unknown potential has a central character and
commutes with the dipole operator, it preserves the simple form
(\ref{simplem1}) for the energy weighted sum rule and the final
numerical results are still compatible with those quoted above.  
A more refined approach to $\alpha$ in the case of the IK model, as
developed within a variational framework \cite{TRAINI94}, also produces 
results which are consistent with the constrains (\ref{cqmestimate}) 
as long as the bounds (\ref{alphabounds}) are considered.

In the following we will investigate boson exchange corrections to the 
simple result (\ref{cqmestimate}) as generated by the inclusion 
of pion and sigma exchanges and to this end it is convenient to discuss 
lower and upper bound separately. In fact they involve different dynamical
effects and ingredients.

\section{Lower Bound: Dipole Sum Rule}

	For the sake of clarity let us begin the investigation of the lower 
bound discussing the analogous nuclear limit, which somehow motivates
our interest in the mesonic exchanges in the interaction. 

When photon-nucleus interactions are considered, it is well
known (see e.g. Ref. \cite{BETHE50}) that the first moment of the the dipole
strength can be written:

\begin{equation}
\label{m1nuclear}
m_1(D_z) = e^2 \frac{N Z}{2 M A} (1 + K)	,
\end{equation}

\noindent
where $N$ and $Z$ represent the number of neutrons and protons
respectively and $A=N+Z$. The first term in Eq. (\ref{m1nuclear}) arises from the
kinetic energy conmutator  and $K$ is an enhancement factor due to the
(isospin--dependent) N-N interaction arising from boson exchange.  
Considering a OPE potential
model the values for $K$ range from 0.5 to 1, depending on the
importance of the tensor correlations \cite{ORLANDINI91}.

The nuclear example makes clear the modifications to $m_1(D_z)$ one should expect
in the case of nucleon. An interesting feature, indeed, distinguishes
the nucleon from the nuclear wave function: due to the color degree of freedom 
the spin-isospin wave function for the (dominant) $SU(6)$ symmetric component
of the three-quark system, must be symmetric whereas it results antisymmetric
in the nuclear case. As a consequence we expect  corrections to the energy weighted 
sum rule (\ref{simplem1}) consistent with

\begin{equation}
m_1(D_z) = \frac{e^2}{3 m_q} (1+ \kappa)
\label{TRKnuc}       ,
\end{equation}
where $\kappa$ embodies the additional contribution coming from 
$V_{\mbox {\scriptsize OPE}}$ (cfr. Eq.(\ref{vope})) and constrained by
$-1 < \kappa \leq 0$ because of the symmetry properties of the quark wave 
function. 
 
	An explicit calculation of this factor $\kappa$ gives:

\begin{equation}
\kappa  =  - \frac{3}{2} m_q 
\langle 0 |(\vec{\tau}_1 \cdot \vec{\tau}_2 - {\tau}_1^z {\tau}_2^z) r_{12}^2 
\left[ V^S_{\mbox {\scriptsize OPE}}(r_{12}) 
\vec{\sigma}_1 \cdot \vec{\sigma}_2 + 
V^T_{\mbox {\scriptsize OPE}}(r_{12})
\hat{S}_{12} \right] |0 \rangle \,\,,
\label{kappadef}
\end{equation}

	By making use of $q-q$ potential model proposed by Valcarce
et al. \cite{FERNANDEZ93} and its numerical solution \cite{PRIVATE}
\footnote{Quite recently the potential model of
ref. \cite{FERNANDEZ93} has been criticized by Glozman {\it et al.}
\cite{GLOZMAN98} which consider the agreement with the empirical
spectrum be only apparent and obtained because of the truncation in
the hypercentral components. However, it has been argued
\cite{FERNANDEZcomm} that the authors of ref. \cite{GLOZMAN98} did
not take into account the dependence of the regularization of the OGE
interaction on the model space used in the calculation and the
agreement can be restored by considering consistent
regularizations. The discussion is still going on
\cite{GLOZMANetalreply}, but, as it will become clear in the
following, our numerical conclusions are quite independent on the
specific debate and the results obtained making use of the criticized
potential are quite similar to other potential models.} for the wave
function one gets, for both proton and neutron, the surprising {\it
positive} value $\kappa=0.088$ which produces a {\it reduction} of the
lower bound of (roughly) $10\%$.  In spite of the change in the
spin-isospin wave function symmetry and contrary to our expectations
$\kappa$ keeps the sign of the nuclear enhancement factor $K$.

	In order to understand this non intuitive result
it is very instructive to investigate more closely the corresponding moment 
$m_1(D_z)$ in nuclei. In addition if we want to compare the results 
obtained in the two systems, the framework must be as consistent as possible.

Let us consider, therefore, the N-N potential generated by the $q-q$ 
interaction we used to calculate the lower bound to $\alpha$, namely
the nuclear potential of ref.\cite{FERNANDEZ93}: the interaction
contains a central part and an isospin dependent term and
it provides a good description of phase shifts and deuteron properties.
By assuming a simple gaussian wave function for a 
three nucleon system with a harmonic oscillator parameter 
$\alpha_{\mbox {\tiny ho}}^2 = 0.5$ fm$^{-2}$, the enhancement 
factor in (\ref{m1nuclear}) becomes $K \approx 0.1$. The spin-isospin matrix 
elements contain the expected opposite sign (with respect to the nucleon case) 
as already anticipated, but the spatial integral contributes with an additional 
opposite sign and both $\kappa$ and $K$ are positive. The underlying
reason is that the nucleon and the nuclear wave functions are 
sensitive to very separate regions of the OPE potential. 
Let us look at the OPE potential model of ref.\cite{FERNANDEZ93}
in more detail. It can be written

\begin{eqnarray}
\label{vopeF1}
V^S_{\mbox {\scriptsize OPE}}(r_{ij}) & = & \frac{1}{3} \alpha_{ch}
\frac{\Lcsb^2}{\Lcsb^2 - m_\pi^2} m_\pi \left[ Y(m_\pi r_{ij}) -
\frac{\Lcsb^3}{m_\pi^3} Y(\Lcsb r_{ij})\right] \\ V^T_{\mbox
{\scriptsize OPE}}(r_{ij}) & = & \frac{1}{3} \alpha_{ch}
\frac{\Lcsb^2}{\Lcsb^2 - m_\pi^2} m_\pi \left[ H(m_\pi r_{ij}) -
\frac{\Lcsb^3}{m_\pi^3} H(\Lcsb r_{ij})\right],
\label{vopeF2}
\end{eqnarray}

\noindent
where
\begin{equation}
\label{endvqq}
Y(x) = \frac{e^{-x}}{x}\,\,,\:\:\: H(x) = \left( 1 + \frac{3}{x} + 
\frac{3}{x^2}\right) Y(x)\,\, ,
\end{equation}
and $m_\pi$, $m_\sigma$ and $m_q$ are the pion, sigma and quark masses 
respectively, $\alpha_{Ch}$ is the chiral coupling constant.
The cut-off parameter $\Lcsb$ controls the size of the pion-quark 
interaction region.

From Eqs. (\ref{vopeF1}-\ref{vopeF2}) 
and taking into account the rather large value 
for the cut-off $\Lcsb$ (4.2 fm$^{-1}$), one can check that for 
distances $\lesssim 1.5$ fm the regularized part of the interaction dominates. 
The nucleon wave function is mostly concentrated within distances 
$\lesssim 0.5$ fm and is sensitive to the short range (or regularized) part, 
while the nuclear wave function is mainly sensitive to the long range tail 
of the OPE.  This simple argument justifies the change of sign previously 
emphasized and we can conclude that, despite of the use of the same terminology,
the only common feature of the OPE in nucleons and in nuclei is 
its spin-isospin structure. The spatial structure of the  
OPE interaction looks very different moving from nucleons to nuclei:
the quarks in the nucleon are sensitive to the short-range part only,
which survives in the chiral limit \cite{GLOZMANpanic} but is poorly
known, whereas the nucleons in the nuclei are mostly influenced by the
long-range tail.

At this point one might wonder whether our conclusion for $\kappa$ can
be considered as general or it is a particular feature of the employed
model. To check this point, and taking advantage of the
flexibility of the sum-rule techniques, we have repeated the
calculation of $\kappa$ for other interquark potential models, namely,
the hybrid model proposed 
by Dziembowski, Fabre de la Ripelle and Miller in ref.\cite{GLOZMAN96} 
and the version of the OBE model due to Glozman, Papp, Plessas, Varga and 
Wagengrunn presented in \cite{GLOZMAN98}. 
For those potentials we used a simple coulombian-type ($\Psi \propto
\exp(-\xi/\beta)$) approximation for the wave functions 
whose size parameter is fixed
to reproduce the proton root-mean-square radius predicted by those 
potential models.
The sum rule value of $m_0(D_z)$ is fixed by the size of the nucleon and therefore
is well reproduced also within such approximate procedure. The other
ingredient of the lower bound, namely $m_1(D_z)$, is quite insensitive to the 
details of the wave function behavior once the size of the system is 
reproduced\cite{ORLANDINI91}. For example, for the potential of
Valcarce {\it et al.} the coulombian approximation of the wave
function gives $\kappa = 0.075$ against $\kappa=0.088$ of the
calculation with the full wave function. 

Our results for $\kappa$ are summarized in table I, where we can see
that in spite of the differences in the absolutes values (due
essentially to the different strength $\alpha_{Ch}$ and cut-off
$\Lambda$ parameters) the sign of $\kappa$ is always positive. Therefore  
our previous conclusions are quite general and rely on the common short range 
behavior of the interquark potentials, as it is confirmed in Fig. 1
where they are compared in the region $r_{12} \lesssim 5$ fm. In fact
these similarities are a direct consequences of the rather large
values of $\Lambda$ employed in the models. In conclusion, the introduction 
of meson exchange contributions lead to a smaller value of the lower bound
and their inclusion does not represent any improvement.

\section{Upper Bound: Two-Body Charge Densities}

	The discussion of the previous section leads to the conclusion that
the lower bound to the nucleon polarizability is rather small even including
meson exchange effects in the $q-q$ potential. Does such conclusion mean
that the range of the allowed values for $\alpha$ is simply enlarged 
by the presence of virtual mesons? The question opens the need for
a reinvestigation of the 
upper bound. In fact it is evident from Eq. (\ref{alphabounds}) that a shift 
in the allowed values of $\alpha$ is achieved more easily  by increasing 
$m_0(D_z)$ rather than by lowering $m_1(D_z)$. In addition
one can note that $m_0(D_z)$ is extremely transparent: it depends on the 
definition of the dipole operator (and the nucleon ground state) only.
The expression for $D_z$ used in the previous section was obtained 
by assuming a non-relativistic charge density:

\begin{equation}
\label{rhonr}
\rho_{\mbox {\tiny NR}}(\vec{q}\:) = \sum_{i=1}^3 e_i e^{i \vec{q}\: (\vec{r}_i - \vec{R})}  ,
\end{equation}

\noindent
the dipole operator being defined as:

\begin{equation}
\label{dzgendef}
D_z = - i \left. \frac{\partial \rho(\vec{q}\:)}{\partial q_z}
\right|_{\vec{q} \rightarrow 0}.
\end{equation}
Corrections to the charge density \cite{FRIAR77} will translate into 
modifications to the dipole operator, i.e., to $m_0(D_z)$. However there are
modifications of the charge density which leave the dipole operator unchanged.

	As an example let us consider the rather common inclusion of a photon-quark 
form factor which is supposed to take into account the structure of the 
constituent quarks:

\begin{equation}
\label{qgammaff}
e_q \rightarrow e_q(q^2) = \frac{e_q}{1 + \frac{1}{6} \vec{q}\:^2 
r_{\gamma q}^2}\,\,.
\end{equation}

\noindent
The replacement (\ref{qgammaff}) increases the charge radius \cite{BUCHMANN94} 
of the baryon but has no effect on the dipole operator. The same conclusion 
holds for any relativistic correction to (\ref{rhonr}), such as the 
Darwin-Foldy term, which does not modify the spherical symmetry of the quark
charge distribution.

	A trivial solution, sometimes invoked, would be the introduction of
{\it ad hoc} intrinsic polarizabilities of the quark (or equivalently,
intrinsic dipole form factors, analogous to (\ref{qgammaff})), but the predictive 
power of the model is lost unless it is extended to the study of other 
observables (for example generalized polarizabilities 
in virtual Compton scattering). A more ambitious approach, 
beyond the scope of the present work, would be to explain how this intrinsic
polarizabilities (or intrinsic form factors) 
are built from more fundamental physical mechanisms,
such as pion loop fluctuations (see for instance \cite{WEINBERG90}).

	Since we are interested in those mesonic effects directly
related to the spectral Hamiltonian we will investigate in detail
a source of corrections to the charge density (and consequently
to the dipole operator of Eq.(\ref{dzgendef})) which is directly related 
to the form of the potentials  we are studying: the two-body exchange terms 
which appear as a consequence of current conservation when isospin or velocity 
dependent interactions are included.
The $q-q$ potentials considered in the present investigation contain, in fact, 
terms of this kind: the OPE interaction (\ref{vope}). 
In addition to the pion exchange current required by current conservation

\begin{equation}
\label{pioncurrent}
\vec{q}\cdot \vec{J}{\mbox {\scriptsize OPE}} = 
\left[V_{\mbox {\scriptsize OPE}}, \rho_{\mbox {\tiny NR}} \right]\,\,,
\end{equation}

\noindent 
a two-body charge density $\rho^{\mbox {\tiny OPE}}(\vec{q}\:)=
\rho_{\pi q \bar{q}}(\vec{q}\:)$
can be associated (see \cite{BUCHMANN97} and references therein)
and the corresponding modification to the dipole operator considered.
Let us emphasize that, in fact, the pion exchange current contributions
(\ref{pioncurrent}) have been already included in the calculation
of the energy weighted sum rule $m_1(D_z)$ of Eq.(\ref{TRKnuc}).
They are embodied in the additional term $\kappa$ which does not vanishes
precisely because the conmutator (\ref{pioncurrent}) is different from zero
\footnote{In photonuclear physics this is known
						     as Siegert theorem.}.
Selfconsistency would, therefore, require to consider also the additional 
mentioned two-body densities in $m_0(D_z)$. 
The order by order procedure can stop here 
(as to the first order pion terms) and we do not 
need to reconsider two-body densities contribution to $m_1(D_z)$.

Let us explicitly show the corrections to the dipole operator
coming from the pion exchange interaction (\ref{vopeF1}). One has

\begin{eqnarray}
D_z^{\mbox {\tiny OPE}} & = & - i \left. 
\frac{\partial \rho^{\mbox {\tiny OPE}}(\vec{q}\:)}{\partial q_z}
\right|_{\vec{q} \rightarrow 0} = \nonumber \\
& = & -e \frac{\alpha_{Ch}}{m_q} 
\frac{\Lcsb^2}{\Lcsb^2 - m_\pi^2}
\sum_{i<j}\left[ \left( \frac{1}{6} \vec{\tau}_i \cdot \vec{\tau}_j + 
\frac{\tau_{j\,z}}{2}
\right) \sigma_{i\,z} \cdot \vec{\sigma}_j \hat{r}_{ij} +
(i \leftrightarrow j)\right] \nonumber \\
& & \left( G(m_\pi r_{ij}) - \frac{\Lcsb^2}{m_\pi^2} G(\Lcsb r_{ij})
\right) \; ,
\label{Dope}
\end{eqnarray}
with 
\begin{equation}
G(x) = \left( 1 + \frac{1}{x} \right) Y(x).
\end{equation} 

From the numerical point of view the pion contribution in the model of
Valcarce {\it et al.} {\it lowers} the
non-energy weighted sum rule from $m_0(D_z)=4.88 \; 10^{-4}$ fm$^2$ to
$m_0(D_z)=4.55 \; 10^{-4}$ fm$^2$ for the proton and 
$m_0(D_z)=4.38 \; 10^{-4}$ fm$^2$ for the neutron
(the neutron - proton difference originates not only 
from the $SU(6)$-breaking components in the nucleon wave function, but
also from the structure of the operator (\ref{Dope}) which is sensitive 
to the total isospin of the system even in the $SU(6)$-symmetric limit).

The modifications in $m_0(D_z)$ renormalize slightly both upper and
lower bounds to the polarizability which becomes

\begin{equation}
\begin{array}{rcl}
\label{OPEbounds}
2.48 \; 10^{-4} \;\mbox{fm}^3 \leq & \alpha^{p} & 
\leq 3.51 \; 10^{-4} \;\mbox{fm}^3 \nonumber \\
2.29 \; 10^{-4}\; \mbox{fm}^3 \leq & \alpha^{n} & 
\leq 3.38 \; 10^{-4}\; \mbox{fm}^3 \,\,.
\end{array} 
\end{equation}

	A similar trend is observed for other potentials (see Table
II). The inclusion of $\rho_{\pi q \bar{q}}$ has a small effect on
$m_0$ and in general it reduces the initial one body contribution. As
a result the  upper bounds to $\alpha^p$ with the other two
potentials shown in Table II are $4.2 \; \cdot 10^{-4}$ fm$^3$ for
Dziembowsky {\it et al.} and $1.49 \; \cdot 10^{-4}$ fm$^3$ for
Glozman {\it et al.} 

	A comparison with Eq.(\ref{simplem0}) shows that the present 
results for $m_0$ are consistent with a shrinking of the size of the
nucleon by OPE, an effect already discussed by the authors of refs.
\cite{BUCHMANN94,BUCHMANN97,HELMINEN99}. Two-body mesonic currents
have been proved to be 
crucial to explain some observables such as $N-\Delta$ electric multipoles 
and neutron charge radius \cite{BUCHMANN97}. However in these cases the 
leading one-body contribution is strongly suppressed by symmetry reasons.  
The argument is not valid in general and cannot be invoked for the electric 
polarizability  (neither for the charge radius of the proton) where these two-body
currents are not sufficient to parameterize all the non-valence degrees 
of freedom seen by electromagnetic probes.
Nevertheless before drawing more definite conclusions on the effects of
meson exchange on the nucleon electric polarizability, it is worth 
mentioning that two-body charge density modifications come also from the other
terms of the $q-q$ interaction. Their origin is a little more subtle than
the pion contributions and has to do basically with the relativistic corrections 
to the one-body charge density of Eq.(\ref{rhonr}). As a result the charge
density can be written as a sum of a non-relativistic contribution plus 
the ones coming from OPE, OSE, OGE and confinement potential; namely
\begin{equation}
\label{rhofull}
\rho(\vec{q}\:) = \rho_{\mbox {\tiny NR}}(\vec{q}\:) + 
\rho_{\pi q \bar{q}}(\vec{q}\:) + 
\rho_{\sigma q \bar{q}}(\vec{q}\:) +
\rho_{g q \bar{q}}(\vec{q}\:) +
 \rho_{\mbox{\tiny conf}}(\vec{q}\:)\,\,.
\end{equation} 
On the same basis the dipole operator acquires the corresponding 
components:
\begin{equation}
\label{dzdecomposition}
D_z = D_z^{\mbox {\tiny NR}} + D_z^{\mbox {\tiny OPE}} + 
D_z^{\mbox {\tiny OSE}} + D_z^{\mbox {\tiny OGE}}+
D_z^{\mbox {\tiny conf}} \, \, .
\end{equation}

Since we want to focus on mesonic contributions, we will not consider
here the OGE and confinement component, and the only additional
component we will calculate explicitly is the one coming form the  
the OSE potential term, obtaining
 
\begin{eqnarray}
\label{Dose}
D_z^{\mbox {\tiny OSE}} & = & - \alpha_{Ch} \frac{1}{2 m_q} 
\left( \frac{m_\sigma}{m_\pi} \right)^2 
\frac{\Lcsb^2}{\Lcsb^2 - m_\sigma^2} \nonumber \\
 & & \sum_{i<j} \left[ \left( \frac{1}{2} e_i r_{i\,z} m_\sigma + 
(i \leftrightarrow j)\right) 
 \left( G(m_\sigma r_{ij}) - \frac{\Lcsb^2}{m_\sigma^2} G(\Lcsb
r_{ij})\right) \right. \nonumber \\
& & \left. + \left( e_i \hat{r}_{ij} + (i \leftrightarrow j) \right)
\left( Y(m_\sigma r_{ij}) - \frac{\Lcsb^3}{m_\sigma^3} 
Y(\Lcsb r_{ij})\right) \right] \; .
\end{eqnarray}

The detailed contributions to $m_0(D_z)$  
 from the different terms of
the dipole operator $D_z= D_z^{\mbox {\tiny NR}} + D_z^{\mbox {\tiny OPE}} + 
D_z^{\mbox {\tiny OSE}}$ are summarized in table III.
The full result, sum of all the entries in the table II, 
is $m_0(D_z)=4.66 \; 10^{-4}$ fm$^2$ for the proton and 
$m_0(D_z)=4.51 \; 10^{-4}$ fm$^2$ for the neutron, values slightly 
larger than the corresponding results (\ref{OPEbounds}) obtained 
including OPE only, but still smaller than non-relativistic 
impulse approximation obtained considering the one-body contribution 
$D_z^{\mbox {\tiny NR}}$ only. Therefore, the final bounds on the
 polarizability remain:

\begin{equation}
\label{finalbounds}
\begin{array}{rcl}
2.60 \; 10^{-4} \;\mbox{fm}^3 \leq & \alpha^{p} & 
\leq 3.59 \; 10^{-4} \;\mbox{fm}^3 \nonumber \\
2.43 \; 10^{-4}\; \mbox{fm}^3 \leq & \alpha^{n} & 
\leq 3.48 \; 10^{-4}\; \mbox{fm}^3 \,\,,
\end{array}
\end{equation} 

A final comment on the terms neglected in (\ref{dzdecomposition}) is
in order. Previous analysis of the contribution of $\rho_{\mbox{\tiny
conf}}$ to the nucleon electric form factors show
a quite large sensitivity to the potential model: in \cite{BUCHMANN97}
the confinement two-body charge produces a reduction of the charge
radius of the proton whereas in \cite{HELMINEN99} an increment is
found. Furthermore, there is some  sensitivity to the
details of the employed wave function (see \cite{BUCHMANN94} for a
comparison). While the $\rho_{\mbox{\tiny conf}}$ term is present in
both OBE-based and hybrid models, in the latter we can find an
additional two-body density $\rho_{g q \bar{q}}(\vec{q}\:)$ that gives
a positive contribution to the size of the proton 
\cite{BUCHMANN94,BUCHMANN97}. 
One can estimate that this increment of the
square charge radius of the proton, that in our reference model is of
0.063 fm$^2$, would presumably induce an increment (through $m_0$) 
of the allowed values for $\alpha^p$
of, roughly, $\lesssim 1 \cdot 10^{-4}$ fm$^3$. Nonetheless, in
some models \cite{BUCHMANN97} such a large effect on the size of the
nucleon is strongly suppressed by the confinement two-body current.
Since the emphasis of our
work is on meson exchange effects we will not discuss this effects further. 

	Additional relativistic corrections to the current could
certainly be considered \cite{DESANCTIS90,FRIAR77,DESANCTIS89}, and
they would also receive contributions from pion and sigma exchange.
In ref. \cite{DESANCTIS90} the effects of relativistic corrections to
$\rho_{\mbox {\tiny NR}}(\vec{q}\:)$ were calculated (up to order
$(1/m^2)$) and found to be small ($\lesssim 5$ \%) and also reduced
the value of $\alpha$.  On the same grounds, potential dependent
corrections to the static polarizability have been proposed
\cite{BERNABEU98} and discussed \cite{LVOV98}.  However we believe
that the comparison between (\ref{finalbounds}) and
(\ref{cqmestimate}) gives a good idea of the importance of the meson
exchange for the nucleon polarizability. In addition the validity of
our conclusions are largely independent on the details of $q-q$
interaction model and rely on their basic features. In particular the
aspects which play a relevant role for the present investigation are
the small core size for the nucleon wave function (which basically
determines $m_0$) and the sign and short-range behavior of the
mesonic interaction (that enters in $m_1$), largely constrained by the
$N-\Delta$ splitting in the spectrum.

\section{Summary}

	We have shown that the nucleon polarizability is
quite a transparent observable to elucidate the r\^ole 
(or better, the lack of r\^ole) of meson-exchange
effects between quarks in a non-relativistic approach. 
The robustness of the employed method, the sum rule technique, resides
on the fact that the sum over infinite excitations is avoided and 
transformed into conmutators that depend on very basic features of the
$q-q$ interaction, so that it is possible to compare, in
a simple and direct way, a large class of  $q-q$ potential models.
It has been shown that, unlike in nuclei, meson
exchange currents are not relevant to explain experimental
evidences. 

	The comparison between the nuclear and the nucleon case reflects the 
rather different scales involved in the problem. While the former is 
sensitive to the long-distance tail of the OPE potential the last is 
mostly affected by the regularized short-range part. A different choice of 
the OPE interaction does not change substantially 
the behavior of $\kappa$ in section III. This is basically a 
consequence of the constraints that hadronic spectroscopy (and in 
particular the $N-\Delta$ splittings) imposes on the short-range 
behavior of the pseudoscalar potential. Furthermore, the small size
of the nucleon seems to be also a common requirement of the baryonic
spectroscopy.
 
	Another type of corrections to the polarizability are due to the two-body 
charge density generated by meson exchanges ($\pi$ and $\sigma$ 
considered here in more detail). Their contribution tends to reduce the allowed 
values of $\alpha$. The same trend is observed in other calculations 
for the charge 
radius of the proton \cite{BUCHMANN97,HELMINEN99}. These facts are pointing 
out the limitations of mesonic two-body terms to parameterize the effects of 
non-valence degrees of freedom in the nucleon. The gluonic two-body
current present in OGE-based and hybrid model might contribute to some
extent to increase the allowed values for the polarizability.
 
	The picture of the nucleon as a core plus a meson cloud,
that emerges not only from constituent quark models but also from
Nambu--Jona-Lasinio and Skirmion models \cite{LUTZ90}, have no analogy
in nuclei. Meson exchange currents, or equivalently the OPE and OSE terms,
still leave much room to the cloud in the 
description of the electromagnetic properties of the nucleon. The
inclusion of the internal polarizability of pions and/or more refined
treatment of the $|qqq \; q\bar{q}\rangle $ component in the wave
function seems to be unavoidable and it should be further investigated.
In other language, it means that the efforts in the description of the
electromagnetic baryonic transitions should be concentrated on the
construction of the effective photon-quark (and/or
photon-quark-antiquark, etc) operator since the hadronic wave
functions cannot account for all the physical ingredients required to
explain the electromagnetic structure of the baryons. 
This conclusion is reinforced by some calculations of other electromagnetic
observables, for which the use of hadronic wave functions based on OBE potentials
\cite{BUCHMANN94,BOFFI99} does not represent much improvement with
respect to the calculations based on OGE interactions \cite{CAPSTICK92}.

\section*{ACKNOWLEDGMENTS}

	We acknowledge illuminating discussions with A. Valcarce, P. Gonz\'alez 
and G. Orlandini. We thank A. Valcarce for providing us with the 
numerical wave functions and useful information concerning 
ref. \cite{FERNANDEZ93}. We also are grateful to A. Buchmann for
his interesting comments on the manuscript.

\newpage

\begin{table}
{\tabcolsep 3pt 
\begin{tabular}{lccc}
\hline 
\rule{0pt}{4.ex}  & Valcarce {\it et al.} \cite{FERNANDEZ93} &
  Dziembowski {\it et al.} \cite{GLOZMAN96}  &
 Glozman {\it et al.} \cite{GLOZMAN98}  \\ 
\hline \rule{0pt}{4.ex}  $\kappa$ & 0.088 & 0.164 (0.143) & 0.139 (0.138) \\ 
\hline
\end{tabular}}
\caption{Comparison between different values of $\kappa$, as defined
in Eq. (\ref{kappadef}), predicted by three potential models. 
Values in parenthesis correspond to a gaussian wave function approximation.}
\end{table}

\begin{table}
{\tabcolsep 3pt 
\begin{tabular}{lrccc}
\hline 
\multicolumn{2}{c}{\rule{0pt}{4.ex}}  & Valcarce {\it et al.} \cite{FERNANDEZ93} &
  Dziembowski {\it et al.} \cite{GLOZMAN96}  &
 Glozman {\it et al.} \cite{GLOZMAN98}  \\ \hline
 $m_0(D_z^{\mbox {\tiny NR}})$ & p,n &
  4.88 & 5.15
 & 2.25 \\ $m_0(D_z^{\mbox {\tiny NR}} + D_z^{\mbox {\tiny OPE}}) $  & 
 p & 4.55 & 5.59 & 2.19 \\
& n  & 4.38  & 4.98 & 1.94  \\
\hline
\end{tabular}}
\caption{Contribution of the OPE two-body charge to the $m_0$ sum rule
(in 10$^{-4}$ fm$^2$ units) for different models. }
\end{table}

\begin{table}
{\tabcolsep 3pt 
\begin{tabular}{lcccccc}
\hline 
\rule{0pt}{5.ex} & $\langle D_z^{\mbox {\tiny NR}}  D_z^{\mbox {\tiny NR}}\rangle $ &
\begin{tabular}{cc}
$\langle D_z^{\mbox {\tiny NR}}  D_z^{\mbox {\tiny OPE}} $ \\ 
$ +  D_z^{\mbox {\tiny OPE}}  D_z^{\mbox {\tiny NR}}\rangle$ 
\end{tabular} &  
\begin{tabular}{cc}
$\langle D_z^{\mbox {\tiny NR}}  D_z^{\mbox {\tiny OSE}} $ \\
$ +  D_z^{\mbox {\tiny OSE}}  D_z^{\mbox {\tiny NR}}\rangle$ 
\end{tabular} &
\begin{tabular}{cc}
$\langle D_z^{\mbox {\tiny OSE}}  D_z^{\mbox {\tiny OPE}} $ \\ 
$ +  D_z^{\mbox {\tiny OPE}}  D_z^{\mbox {\tiny OSE}}\rangle$ 
\end{tabular} &
$\langle D_z^{\mbox {\tiny OPE}}  D_z^{\mbox {\tiny OPE}} \rangle$ &
$\langle D_z^{\mbox {\tiny OSE}}  D_z^{\mbox {\tiny OSE}} \rangle$ \\[2.ex]
\hline
\rule{0pt}{4.ex} Proton & 4.88 & -1.16 & -0.105 & -0.398 & 
0.831 & 0.620 \\[1.ex]
Neutron & 4.88 & -1.12 & -0.050 & -0.398 & 0.591 & 0.617 \\[3.ex]
\hline 
\end{tabular}}
\caption{Contributions to $m_0(D_z)$ (in 10$^{-4}$ fm$^2$ units)
when two-body OPE and OSE terms are considered in the dipole operator
for the potential model of Valcarce {\it et al.} \protect\cite{FERNANDEZ93}. Each
column shows the contribution of different pieces of the dipole
operator according to Eq.(\ref{dzdecomposition}) with the notation
$\langle {\cal{O}} \rangle = \langle 0 | {\cal{O}} | 0 \rangle$.} 
\end{table}

\newpage
\begin{figure}
\centerline{\protect\hbox{\psfig{file=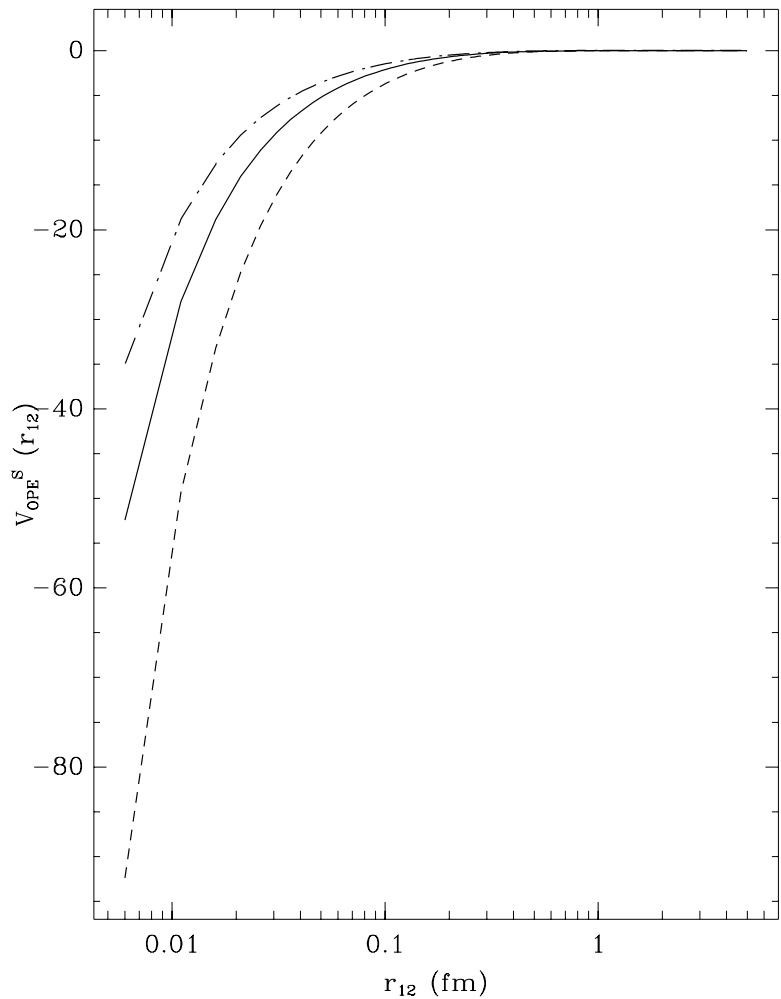,width=0.55\textwidth}}}
\caption{Behavior of the $V^S_{\mbox {\scriptsize OPE}}(r_{12})$ (in
fm$^{-1}$) for different models (solid line: Valcarce {\it et al.} 
\protect\cite{FERNANDEZ93};  dashed line: 
Dziembowski {\it et al.} \protect\cite{GLOZMAN96}; dot-dashed line:
Glozman {\it et al.} \protect\cite{GLOZMAN98}.}
\end{figure}

\end{document}